\documentclass[a4paper,superscriptaddress,floatfix,nofootinbib,onecolumn,longbibliography,notitlepage]{revtex4-1}
\usepackage{bbm}
\usepackage{bbold}
\usepackage{bbm}
\usepackage[pdftex]{graphicx}
\usepackage{latexsym,amsmath,verbatim,amssymb,txfonts}
\usepackage{lipsum}
\usepackage{color}
\usepackage{rotating}
\usepackage{verbatim}
\usepackage{multirow}
\usepackage{hyperref}
\usepackage[english]{babel}
\usepackage{comment}
\usepackage{bm}
\usepackage{amsmath}
\usepackage{amsfonts}
\usepackage{amssymb}
\usepackage{float}
\usepackage{color}
\usepackage{braket}
\usepackage{blkarray, bigstrut}

\begin{document}
\title{What can PhD students and postdocs do to counter inequalities?}

\author{Lorena Ballesteros Ferraz} 
\email{lorena.ballesteros-ferraz@cyu.fr}
\affiliation{LPTC, Université de Lorraine, CNRS, F-54000 Nancy, France \& \\ 
Laboratoire de Physique Théorique et Modélisation, CY Cergy Paris Université, 95302 Cergy-Pontoise, France}

\author{Carolina Charalambous}
\email{caritocharalambous@gmail.com}
\affiliation{Institute of Astrophysics, Pontificia Universidad Católica de Chile, 8331150 Santiago, Chile}

\author{Sébastien Mouchet}
\email{sebastien.mouchet@umons.ac.be}
\affiliation{Research Institute for Materials Science and Engineering, Université de Mons, 7000 Mons, Belgium \& \\ Department of Physics, NISM \& ILEE, University of Namur, 5000 Namur, Belgium \& \\  Department of Physics and Astronomy, University of Exeter, Exeter EX4 4QL, United Kingdom}

\author{Riccardo Muolo}
\email{muolo.r.aa@m.titech.ac.jp}
\affiliation{Department of Systems and Control Engineering, Institute of Science Tokyo (former Tokyo Tech), Tokyo 152-8550, Japan.}

\begin{abstract}
    In this opinion article, we gathered some reflections and practical tips on what Early Stage Researchers do against inequalities in academia. This is the longer version of an opinion paper that was recently published on the website of the International Society for Optics and Photonics (spie.org), containing more details and suggestions. 
\end{abstract}

\maketitle

From the outside, research might seem like a realm where everybody is assessed solely based on their merits, free from discrimination. However, from the inside, we notice every day that this is not the case. It is not only a problem of academia itself, but also the result of living in a society full of inequalities. Being part of society, academia naturally reflects such negative aspects. Can we do something about it?
We are a collective of four Early-Stage Researchers (ESRs) with diverse origins - Argentina, Belgium, Italy and Spain - each hailing from backgrounds in Mathematics, Astronomy, and Physics, who serendipitously converged in time and space within the walls of the University of Namur, in southern Belgium. Our shared fervor for fostering a more inclusive and diverse scientific future catalyzed our involvement in various collaborative projects, to try to answer the above question. Among our endeavors, some of us actively contributed to organizing events celebrating the International Day of Women and Girls in Science and worked towards enhancing the inclusivity of the names in our university public space; as it stands, there isn't a single public space within the university named in recognition of a woman's exceptional accomplishments. The outcomes of these projects were mixed, but one of our primary goals was achieved – the university's debate had been ignited and became a more frequent topic of discussion following our initiative. While Principal Investigators (PIs) and decision-makers possess the capacity to effect substantial changes, we, as ESRs, may feel powerless given our lack of job security and influence within our institutions. However, we have the power of taking small steps towards a more equal future. With this realization in mind, we have chosen to write this column with some of our ideas to improve the academic future from our position, suggesting small actions that, even from our position, we can implement against inequalities. Furthermore, we aspire for this column to kindle a vibrant debate within  the academic community.\\

\paragraph{Raising awareness} One of the most impactful ways for young researchers to address inequalities in academia is by raising awareness. This can include suggesting changes to the wording used in job offers or engaging in informal discussions in coffee rooms and corridors. In Western countries, there is a significant representation of middle-aged white male academics who may not have personal experiences with discrimination, which could potentially limit their understanding of the extent of these inequalities. Nevertheless, some of them may be receptive to the idea of supporting concrete changes. Another effective way to reach fellow researchers is by publishing opinion letters and articles that highlight discriminatory practices, and offer practical solutions. Young researchers are often more attuned to these issues, and it is essential that ESRs take the time to share their insights and experiences to make academia more equitable and inclusive.\\

\paragraph{Supporting students from diverse social backgrounds} It is widely acknowledged that university degrees predominantly cater to individuals from socially privileged backgrounds, especially those with higher levels of education. Hence, it is imperative to ensure that also individuals from different social backgrounds receive adequate financial support to pursue further academic milestones, such as a PhD. Although ESRs may not typically possess the means to provide direct funding, they can still make an impact through small yet significant gestures. These include sharing comprehensive information about available grants or university courses that offer a salary, highlighting university services specifically designed to aid financially disadvantaged students, or providing emotional support. In essence, ESRs can adopt the role of compassionate mentors, ready to assist and guide students, especially from underprivileged backgrounds.\\

\paragraph{Tackling harassment}  Before analyzing someone else's behavior, we need to stop and think individually about our own actions. Even if we believe that we are not doing anything wrong, there is a possibility that we are making someone else feel uncomfortable around us, so we might want to reconsider our own practices and be open to receiving comments or criticisms regarding our attitudes. Tackling harassment does not have a universal solution, but ESRs must commit to preventing harassment and work on prevention by taking some initiatives. For example, having distinctive door tags to signal approachability can be useful in such situations, providing a safe person to talk to. ESRs can also advocate for the presence of neutral parties like gender commissions with specialized workers and community members, working together to ensure that professors and senior staff understand their responsibility to stop, address, and prevent harassment to avoid reaching stressful situations. Such neutral parties can hold regular meetings to discuss these issues, often presented as microaggressions. They can also collaborate with parity movements and relevant associations, creating a support system within the workplace. It might also be beneficial to establish and promote guidelines or codes of conduct to clearly define acceptable behavior within the work environment, helping to reduce harassment, and having a concise and specific anti-harassment policy and up-to-date training procedures. Regular equality and diversity training sessions for all community members can make it easier to recognize when a situation is not right and better equip individuals to stand against bullying and know how to act.\\

\paragraph{Acknowledging the Impostor Syndrome} Members of some underrepresented groups are often reported as more likely to be affected by impostor syndrome, in which individuals doubt their skills and fear being exposed as frauds despite their accomplishments. This syndrome may lead to anxiety, lack of self-confidence, or depression. Despite some efforts, academia tends to largely ignore and neglect impostor syndrome, much like most mental health issues. As ESRs, we have the possibility to raise awareness by discussing psychological wellness challenges with colleagues, tactfully encouraging students and junior colleagues, and organizing seminars and workshops on the subject through our respective institutions' career development programs.\\

\paragraph{Organizing inclusive conferences} In general, conferences and workshops are costly, and the number of scholarships available for underfunded researchers is limited. It is common, especially for women, to wonder how to balance caring for young children or other family members while attending conferences abroad. Many ESRs are involved in organizing conferences and workshops. They may have some power in making such events more inclusive within the limits of the facilities of the venue, the time, and the budget. One potential solution is to provide grants to help cover the costs of childcare and care for family members back home. Alternatively, offering on-site childcare facilities and/or a breastfeeding room could enable more parents to bring their children, allowing them to attend and gain the visibility that they deserve, specifically from women.
Another example is the timelines in conference deadlines. As they are often unconsciously calibrated for Western researchers, they do not usually take into account the time for visa applications. Usually they do not allow enough time between acceptance notification and the beginning of the conference, excluding international attendees and limiting multicultural academic exchange. Similarly, payment methods like credit cards or bank transfers are easy for those in countries with strong and widely used currencies, while they can be nearly impossible for others. While ESRs may not have the power to change certain core aspects that are part of a broader problem, enabling researchers from specific countries to pay registration fees in cash and extending the time frame from acceptance to the start of the conference can make a significant difference. Another important consideration when organizing conferences is ensuring that disabled people can attend without impediments. For example, the venue should have step-free access for people with limited mobility, slide text should be visible, speakers should be asked to speak clearly and not cover their mouths, allowing lip reading for the deaf in the audience. Lastly, organizers may provide some sensory suppression devices and a resting room for neurodiverse people.\\

\paragraph{Avoiding non-inclusive committees} Declining participation  in committees that you estimate lack diversity, with members at different career stages, backgrounds, and genders, while  explaining your reasons is a critical step in creating a more inclusive environment. Such refusals can encourage committees to place greater emphasis on diversity and equality. Only by bringing together diverse voices, we can create a truly inclusive and effective scientific community. It is essential to note, however, that in many institutions or fields of research, the number of female permanent staff is not proportionate to that of males. Striving for gender balance in committees, both for educational and bureaucratic matters, can often lead to an excessive workload placed on the few available female professors. It is imperative to emphasize that achieving gender balance within evaluation committees requires universities to hire more women to ensure  a fair and equitable distribution of tasks.\\

\paragraph{Citing equitably} Another context where inequalities emerge is in scientific literature. Scholars tend to cite papers from researchers whom they already know, often from Western universities, where it is easier to make their work known through conference presentations, social media attention, outreach magazines, and collaborations. It has been shown that researchers from minorities, underrepresented groups and non-Western countries are cited less frequently. Their work does not receive adequate consideration. Despite the thousands of new articles released daily, ESRs can contribute to breaking this cycle with small effort. For example, they can search for conferences in their field held outside Europe and the US, as well as look for local speakers and their relevant papers. ESRs can also try collaborating with groups in developing countries conducting similar research, so that they can share their literature. Alternatively, ESRs can approach researchers from such countries at conferences and inquire about relevant papers.\\

\paragraph{Balancing recommendations} On many occasions, we are asked to provide the names of scholars who are experts in our field, either for paper reviews or committees. It is common to have a list of researchers that we know well, often the ones that our supervisors frequently suggest, typically Western men. With little effort, we can break this cycle by diversifying our list, giving women and researchers from other underrepresented categories a chance to contribute with their insights to paper reviews, committees, or conferences. Making the effort to explore researchers beyond our usual circles can also lead to a more inclusive bibliography, allowing for possible new collaborations.\\

\paragraph{Advocating for accessible publications} For researchers in developing countries, it is not just challenging to publish in high-impact journals but also to access them due to high fees. Many papers are not open access, and the costs are too burdensome for some universities. We can make our publications accessible by using open-access preprint servers like arXiv, bioRxiv, \& ChemRxiv, and update the preprints after the publication process. Additionally, we can enhance the accessibility of our publications by using color-blind-friendly figures with appropriate color scales and thorough descriptions in the captions.\\

\paragraph{Conclusions} With this brief column, we want to draw attention to discrimination, racism, inequalities and other issues rooted in the academic environment. We acknowledge that change takes time, and such problems will not disappear overnight. We are aware that many of the issues that we mention throughout the text extend beyond the purely academic sphere and are problems that are part of the society in which we live. Nevertheless, even though we recognize that injustices exist in many other contexts, we believe that it is necessary to attempt at least to change our own environment. For instance, let us consider the representation of LGBTQIA+ people in academia and contemplate whether an inclusive and safe environment is actually fostered for them. Our intention is not to provide a solution or a definitive list of actions to solve these problems, but rather to start a debate and show that, regardless of the academic role, everyone can take action. We encourage members of the academic community to reach out to us and continue this discussion. We hope that concrete actions will emerge and spread to address these matters. We, as individuals, cannot transform the entire academic environment today, but we can take impactful steps to pave the way for future generations.\\

\paragraph*{Acknowledgements} We are grateful to all our colleagues with whom we have discussed such issues over the past years and to all those who took the time to read the draft and come back to us with precious feedback.

\end{document}